# Performance Evaluation of Two-Stage Shared FDL Optical Packet Switch using Contention Resolution Scheme with Packet Releasing Priority

Ahmed Galib Reza, S.C.Tan, and F.M. Abbou

**Abstract**—This paper proposes a two-stage optical packet switch with second stage of recirculate switch of FDL to reduce the number of the FDL used in the switch for contention resolution. The contention resolution scheme with priority in packet releasing from FDL is tested in the two-stage switch for performance evaluation. Simulation result shows that zero packet loss rate achievable with < 0.8 for 32x 32 two-stage switch.

——————————— ◆ ———————————

## 1 INTRODUCTION

THE next generation communication network is demanding more and more network capacity. Optical packet switching technology is able to deliver the enormous bandwidth of WDM networks. Besides higher bandwidth, it offers high speed, data rate and format transparency, and configurability. One of the major drawbacks in optical packet switching network is contention [1] and therefore, various techniques have been introduced to resolve contention: wavelength conversion, deflection routing and optical buffering which is usually implemented using fiber delay line (FDL) [1][2]. Optical buffers are either single stage, which consists of only one block of delay lines, or multistage which consists of several blocks of delay lines cascaded together, where each block contains a set of parallel delay lines. Optical buffers can also be classified into feed-forward, feedback, and hybrid architectures [2].

For multistage feed-forward buffering, several node architectures applying cascaded 2x2 switching elements which contain optical buffer is proposed [3]. A larger switch fabric can be constructed by cascading a number of these 2x2 elements in, for example, Banyan configuration. Complexity of synchronization is one of the major drawbacks of the proposed approach. SLOB is another kind of optical switch with large optical buffer [4]. Optical packet switching fabric CIOQ-OPS is proposed combin ing both input buffering and output buffering [5]. The problem of Head-Of-Line (HOL) blocking of input buffering is solved using OQ-OPS. Due to input buffering the proposed node may suffer from increased delay and additional cost. In [ref], a multi-stage based on tree structure packet switch is proposed and the performance of a multi-stage buffer is strongly affected by the small number of FDLs in the first stage.

Contention resolution schemes are key determinants of packet-loss performance in any packet switching paradigm. A non-reservation scheduling algorithm is proposed for single-stage shared-FDL switch which fails to guarantee that the cells can get the desired output-port after coming out of FDL [6]. A sequential FDL assignment (SEFA) algorithm is proposed to resolve contention [7]. The proposed algorithm achieves lower packet loss rate for the sake of very high time complexity [8]. Ant colony optimization which constitutes some meta-heuristic is applied to resolve contention resolution in [9]. However, meta-heuristic approach is unable to guarantee the proximity of their solutions to the optimal solution and it produces poor result very often because they converge to local optimum solutions that are far from the optimal one [10].

This paper proposes a two stage optical packet switch to resolve contention. The proposed node provides large buffering capacity to buffer optical packet by implementing FDLs in two stages in auxiliary switch which is simple and low cost. It is difference from [11] as the second stage switch with the feebback optical buffers with recirculated FDLs to reduce the number of FDLs used in the switch for the sake of cost. In order to reduce implementation cost only few FDLs (two FDLs) are implemented in the second auxiliary switch. The rest of the paper is organized as

————————————————
- *A.Galib Reza is with the the Faculty of Information Technology. Multimedia University.*
- *S.C. Tan is with the Faculty of Information Technology. Multimedia University.*
- *F.M. Abbou is with the Al-Madinah International University, Malaysia.*

follows. The proposed node structure is described in Section 2. The contention resolution scheme with packet releasing priority is demonstrated in Section 3. Section 4 evaluates the performance of the proposed node. Finally, Section 5 concludes this paper.

## 2 TWO-STAGE SHARED FIBER DELAY LINE OPTICAL PACKET SWITCH

The proposed node consists of two stages of delay lines illustrated in Fig. 1. It combines two types of switches: main switch, and two auxiliary switches. The main switch is a typical node used in synchronous optical packet switching (OPS) network. The objective of using auxiliary switch is that it is a simple cross-connect. The main switch and the auxiliary switches are controlled by the same control unit which is implemented in the main switch. Therefore, the auxiliary switch does not need any processing power or control unit. In the proposed node, FDLs are implemented in the auxiliary switch in order to use variable number of FDLs.

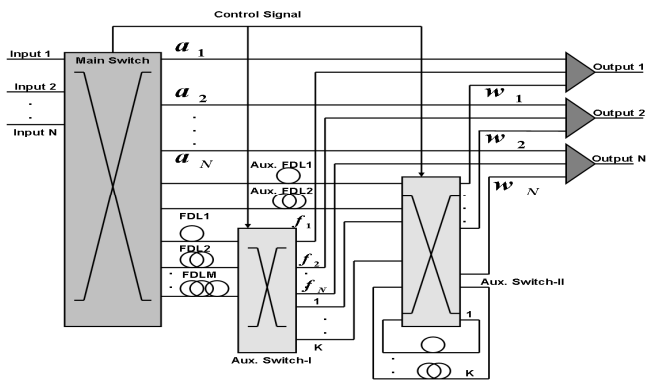

Fig. 1: Proposed two stages shared fiber delay lines optical packet switch.

In this paper, auxiliary switches are denoted as Aux. Switch-I, which contains feed-forward shared FDLs, and Aux. Switch-II, which consists of both feed-forward and feedback shared FDLs. In order to reduce implementation cost, only two FDLs are implemented in Aux. Switch-II for each type of optical buffer. The main switch of the proposed node consists of $N$ input fibers and $(N+M+2)$ output fibers. Among them, $M$ output fibers of the main switch are connected to Aux. Switch-I through fiber delay lines (FDLs) as input. Aux. Switch-I also has $(N+K)$ number of output fibers. The rest of the two output fibers of the main switch are connected to Aux. Switch-II through FDLs and $K$ output fibers of Aux. Switch-I are connected to Aux. Switch-II as input. The Aux. Switch-II also contains $K$ number of feedback FDLs that are shared among all input ports. It consists of $K+K+2$ number of inlets, and $K+N$ number of outlets. Each FDL can delay cells by a fixed number of time slots, and any two FDLs may have the same or different delay values. Each output fiber of the main switch is connected to its corresponding output fiber of both Aux. Switch-I and Aux. Switch-II to produce a single output fiber.

## 3 CONTENTION RESOLUTION SCHEME WITH PACKET RELEASING PRIORITY

The scheduling stage of the proposed node is divided into two parts: Part I. Assigning of FDL, and Part II. Releasing of FDL. It is assumed that each packet is equal in length. The step by step discussion below describes the contention resolution scheme, where *AuxI_FDL* represents the delay lines of Aux. Switch-I, *AuxII_FwdFDL*, and *AuxII_BckFDL* represent the feed-forward and feedback shared FDLs of Aux. Switch-II respectively, and $m$ denotes the number of FLDs of Aux. Switch-I. Take note that Part II of the contention resolution scheme gets higher priority than Part I.

*Part I. Assigning of FDL*

At first, for an incoming packet, if the desired output port is free then it is directly sent to output-port of the main switch without any delay. However, if the output-port is busy then contention occurs and the packet needs to be buffered. For buffering, the contended packet is sent to the free FDL of the Aux. Switch-I in a way that, no two contended packets scheduled at same time destined for same output-port which is similar to the approach of Z. Haas [12]. If such FDL is not found then contended packet is scheduled to send one of the free feed-forward shared FDLs of the Aux. Switch-II which consists of minimum amount of delay. If both of them are busy then contended packet is dropped immediately. This paper difference from [12] as there is packet releasing priority consideration of the switches. The packet stored in the Aux. Switch-II get higher priority in releasing packet compared to Aux. Switch-I.

The basic assignment algorithm when packets are contending for an output port is as follows:

*Step 1:*
*PacketStoredtoAux_I=0;*
*for k = 1 : m*
   *Repeat (for each FDL of Aux. Switch-I)*
   *if (AuxI_FDL[k] is free and no other contended*
      *packet releases from other FDL desiring the same destination at the same time) then*
      *store the contended packet to AuxI_FDL[k];*
      *PacketStoredtoAux_I =1;*
      *break;*
   *end*
*end*
*Step 2:*
*if (PacketStoredtoAux_I == 0) then*
   *for k = 1 : 2*
   *Repeat (for each feed-forward shared FDL of Aux. Switch-II)*
      *if (AuxII_FwdFDL[k] is free) then*
      *store the contended packet to AuxII_FwdFDL[k];*
      *break;*
   *end*
   *end*
*end*
*Step 3:*
*if (packet is not stored in step 1 and step 2) then*
   *drop the packet;*
*end*

*Part II. Releasing of FDL*

In case of releasing optical packets, the packets stored in the Aux. Switch-II get higher priority than Aux. Switch-I.

At first, the packets stored in the feed-forward shared FDLs of Aux. Switch-II that have already elapsed their assigned delay time are sent to their desired output ports. After that, the packets resided in the feedback FDL of Aux. Switch-II are destined for their destination output port. Packets are recirculated in case of busy output port. After successful processing of all the packets of Aux. Switch-II, the contended packets of Aux. Switch-I are scheduled for their destination output port. However, if the output port is busy then packet is sent to one of the free feedback FDLs of the Aux. Switch-II which consists of minimum amount of delay. The packet is dropped instantly if both of the FDLs are found busy.

The basic algorithm for the releasing of contended packets from FDL for the destination output port is given below:

***Step 1***
for k = 1 : 2
   Repeat (for each feedback FDL of Aux. Switch-II)
   if (Destination output port is free) then
      Send the packet to the output port;
   else
      Re-circulate the packet;
   end
end
***Step 2***
for k = 1 : m
   Repeat (for each FDL of Aux. Switch-I)
   if (Destination output port is free) then
      Send the packet to the output port;
   else
      for k = 1 : 2
         Repeat (for each feedback FDL of Aux. Switch-II)
         if (AuxII_BckFDL[k] is free) then
            store the contended packet to Aux-II_BckFDL[k];
            break;
         end
      end
   end
end

## 4 NUMERICAL RESULT AND DISCUSSIONS

In this section, the performance of the contention resolution scheme with packet releasing priority is evaluated in the proposed two-stage 32 × 32 switch. The packet arrives according to Poisson process with rate $\lambda_n$. Traffic is uniformly distributed to all switch output port and it is assumed that packet length is one time unit. Maximum five circulations are considered for two feedback FDLs of Aux. Switch-II and the length of the FDL is linearly distributed as 1, 2, 3, 4… T, where T is the maximum delay value of a FDL and $Z = \log_2 M + 1$ [8], where M denotes total number of FDLs of any switch and Z represents the number of possible delay values for FDLs. For example, when M = 8, then Z = 4. So, if there are 8 optical delay lines of a switch then there will be 4 distinct delay values for FDLs and each value is considered for 2 FDLs.

Figure 2(a) determines the minimum number of FDLS necessary to attain a given packet loss rate under different traffic loads. Three types of traffic load are considered: Low (ρ = 0.3), Moderate (ρ = 0.6), and Heavy traffic load (ρ = 0.9). From Figure 2(a), it is found that under low and moderate traffic load (ρ = 0.3 and 0.6), packet loss rate is zero (PLR = 0) when the number FDLs is equal to 12 and 40 respectively (m = 12 and 40). It is noted that, under heavy traffic load (ρ = 0.9), the number of FDLs are required to increase to 60 (m = 60) to get almost zero packet loss rate (PLR = ~0).

In Figure 2(b), the performance of the proposed switch is illustrated in terms of packet loss rate versus traffic load. It is observed that packet loss rate increases with the increment of traffic load. It is because when traffic load is heavy too many packets arrive in any time slot. As a result, more contention occurs and more packets need to buffer. In Figure 2(b), under heavy traffic load (ρ = 0.9), it is found that the packet loss rate is between $10^{-1}$ and $10^{-2}$ using 32 FDLs (m = 32). The increment of number of delay lines increases buffering capacity thus, reduces the packet loss rate. The improvement of the performance is observed when the number of FDLs is increased to 64 (m = 64). Using 64 FDLs, the packet loss rate is between $10^{-3}$ and $10^{-4}$ which is an improvement of 99.28% of the previous performance.

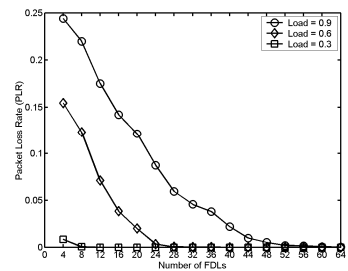

(a)

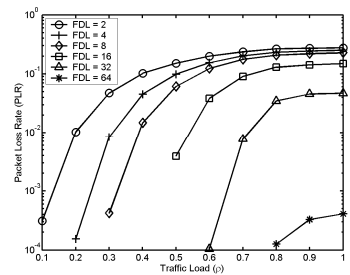

(b)

Fig. 2: Performance of the proposed two stage shared fiber delay line optical packet switch; (a) Packet loss rate vs. number of FDLs, (b) Packet loss rate vs. traffic load.

Figure 3(a) illustrates the performance of the proposed switch in terms of average delay with respect to traffic load (ρ) under different number of FDLs (m). It is observed that when traffic load is low, average delay decreases with the increment of number of FDLs. It is because when the number of FDLs in Aux. Switch-I is less, more packets pass through the FDLs of Aux. Switch-II than in the situation of more number of FDLs in Aux. Switch-I. This may increase the average delay. However, average delay is lower for less number of FDLs than more number of FDLs under heavy traffic load. This is due to the increment of packet loss rate under heavy traffic load. Simulation result shows that under heavy traffic load (ρ = 0.9) average delay is 2.38 time unit for 64 FDLs (m = 64).

Figure 3(b) demonstrates the significance of using Aux.

Switch-II, where *x*-axis represents traffic load ($\rho$) and *y*-axis represents packet loss rate in terms of percentage. A 32 × 32 switch consisting of 32 FDLs (*m* = 32) is used to carry out the simulation. Black shaded area represents the performance of the switch in terms of percentage of packet loss rate without using Aux. Switch-II and gray shaded area represents the percentage of packet loss rate might be reduced if Aux. Switch-II is added on. From Figure 3, it is found that, the packet loss rate can be reduced up to 90% when traffic load is moderate ($\rho$ = 0.6). Under heavy traffic load ($\rho$ = 0.9), Aux. Switch-II is able to reduce the packet loss rate up to 25%. Therefore, it can be concluded that Aux. Switch-II plays a significant role in the reduction of packet loss to improve performance.

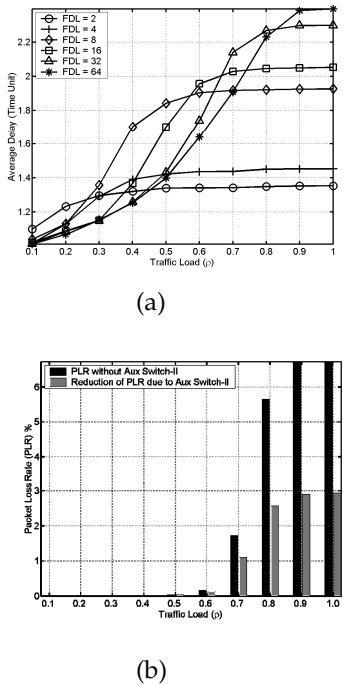

Fig. 3: (a) Average delay vs. traffic load, (b) The Role of Aux. Switch-II in the proposed switch architecture

## 5 CONCLUSIOS

In this paper, a new architectural concept for two-stage optical packet switch to buffer optical packet has been introduced. The proposed switch using very simple two auxiliary switches with second stage of recirculate switch of FDL to resuce the number of the FDL used in the switch. The contention resolution shceme with priority in packet releasing from FDL is evaluated using the proposed switch for performance evaluation. It is shown by the simulation that the packet loss rate of the proposed node is between $10^{-3}$ and $10^{-4}$ at heavy load ($\rho$ = 0.9) for a 32×32 switch using 64 FDLs of different length. It is also shown that the proposed architecture can achieve zero packet loss rate (PLR = 0) when $\rho$ < 0.8.

## ACKNOWLEDGEMENT

This research work is supported by E-Science (No: 01-02-01-SF0089), Ministry of Science, Technology and Innovation (MOSTI), Malaysia